\documentclass[a4paper,11pt]{article}
\pdfoutput=1 

\usepackage{jheppub}
\usepackage{bigints}
\usepackage[T1]{fontenc} 

\title{Butterflies with rotation and charge}
\author{Alan P. Reynolds}
\author{and  Simon F. Ross}
\affiliation{Centre for Particle Theory, Department of Mathematical Sciences,\\Durham University,\\Lower Mountjoy,\\Stockton Road,\\Durham,\\England,\\DH1 3LE}
\emailAdd{a.p.reynolds@durham.ac.uk}
\emailAdd{s.f.ross@durham.ac.uk}

\abstract{We explore the butterfly effect for black holes with rotation or charge. We perturb rotating BTZ and charged black holes in 2+1 dimensions by adding a small perturbation on one asymptotic region, described by a shock wave in the spacetime, and explore the effect of this shock wave on the length of geodesics through the wormhole and hence on correlation functions. We find the effect of the perturbation grows exponentially at a rate controlled by the temperature; dependence on the angular momentum or charge does not appear explicitly. We comment on issues affecting the extension to higher-dimensional charged black holes.}

\begin{document} 
\maketitle
\flushbottom

\section{Introduction}

Holographic studies have recently led to interesting new insights into quantum chaos and the behaviour of entanglement in near-thermal systems \cite{Shenker:2013pqa,Shenker:2013yza,Maldacena:2015waa}. These investigations consider the thermofield double state, 
\begin{equation} \label{tfd} 
|\psi \rangle = \frac{1}{\sqrt{Z}} \sum_i e^{-\beta E_i/2} |E_i \rangle_1 \otimes |E_i \rangle_2,
\end{equation}
which describes an entangled state between two quantum systems with isomorphic Hilbert spaces $\mathcal H_1$, $\mathcal H_2$, where $|E_i \rangle$ are the energy eigenstates, and $Z = \mbox{tr}\, e^{-\beta H}$ is the partition function, included for normalization. Tracing over one copy leads to a thermal density matrix in the other. When we consider this state in a conformal field theory with a holographic dual \cite{Maldacena:1997re}, it can be described holographically by an eternal black hole, with the two copies of the Hilbert space identified with the two asymptotic boundaries of the black hole \cite{Maldacena:2001kr}. This state purifies the thermal density matrix by a specific pattern of short-range entanglement between the two copies of the CFT; this creates non-zero correlations between operators in the two copies, $\langle O_1 O_2 \rangle \neq 0$. The thermal density matrix is time-independent, which is reflected by the invariance of \eqref{tfd} under time evolution with $H_1 - H_2$. However, if we apply evolution with $H_1 + H_2$, the state evolves in a non-trivial fashion, with the entanglement between the two copies becoming more non-local (as signalled by a decay of the two-sided correlators of local operators). This time-evolution of the entanglement was studied holographically in \cite{Hartman:2013qma}, finding that the entanglement spreads out to larger distance scales linearly in time.  

In \cite{Shenker:2013pqa}, Shenker and Stanford initiated a study of perturbations of the thermofield double state, to study the behaviour of near-thermal systems.\footnote{The investigation of perturbations of the thermofield double state is also motivated by the conjecture that generic entangled states are related to wormholes (ER=EPR) \cite{Maldacena:2013xja}; see also \cite{Marolf:2013dba,Balasubramanian:2014gla}.} They considered a small perturbation $W$ added to one of the two CFTs at some early time $-t_w$, and studied its effect on the structure of the state at $t=0$.\footnote{Initially the perturbation was taken to be spatially homogeneous, although localised perturbations were later considered in \cite{Roberts:2014isa}. We will restrict attention to homogeneous perturbations.} As we will review below, for large $t_w$, the perturbation can be modelled by a shock wave near the horizon of the black hole. They considered specifically two-dimensional conformal field theories, for which the dual black hole geometry in the thermofield double state is the non-rotating BTZ black hole. The perturbation deforms the geometry of the wormhole connecting the two asymptotic regions at $t=0$, with the length of a geodesic connecting the two boundaries in the perturbed geometry being given by 
\begin{equation}
\frac{d}{l} = 2\log\frac{r}{r_+} + 2\log\left(1 + \frac{\alpha}{2}\right), 
\end{equation}
where $l$ is the AdS scale, $r$ is a large-distance cutoff, $r_+$ is the radius of the black hole horizon, and 
\begin{equation}
\alpha \sim \frac{E}{M} e^{\frac{2\pi}{\beta} t_w}
\end{equation}
is a parameter controlling the strength of the shock. 

This growth in the length of the geodesic is reflected in the correlation functions for generic local operators in the two CFTs; for a given operator $V$ of dimension $\Delta$, we can approximate 
\begin{equation} \label{ggrowth}
\langle W | V_L V_R | W \rangle \sim e^{-\Delta d/l}  \sim r^{-2\Delta} \left(1 + \frac{\alpha}{2}\right)^{-2\Delta}. 
\end{equation}
 We see that the effect of the early perturbation on the correlation function at $t=0$ grows exponentially in $t_w$; this is a sign of sensitive dependence on initial conditions. The exponential growth produces an appreciable effect on the correlator when $\alpha$ becomes of order one, at the scrambling time \cite{Sekino:2008he}. In \cite{Maldacena:2015waa}, the value of the commutator $C(t) = - \langle [W(t), V(0)]^2 \rangle$ was adopted as a signature of quantum chaos. The behaviour of the commutator is controlled by the out of time order (OTO) correlation function $\langle V(0) W(t) V(0) W(t) \rangle$, which can be related to \eqref{ggrowth} as will be reviewed below.  The lengthening of the geodesic can also be related to changes in the entanglement structure of the dual state through the Ryu-Takayanagi proposal \cite{Ryu:2006bv}.

The behaviour found in \cite{Shenker:2013pqa} is understood to be robust and generic in the space of theories. The calculations were extended to multiple shocks in \cite{Shenker:2013yza}, to localised shocks in \cite{Roberts:2014isa}, and to include stringy corrections in \cite{Shenker:2014cwa}. Field theory arguments have shown that these results apply  not just to CFTs with a holographic dual, but much more generally \cite{Roberts:2014ifa,Maldacena:2015waa,Berenstein:2015yxu,Gur-Ari:2015rcq,Stanford:2015owe}. 

Another natural extension is to consider the behaviour in the presence of chemical potentials for charge or angular momentum, where the thermofield double state is generalised to 
\begin{equation} \label{ctfd} 
|\psi \rangle = \frac{1}{\sqrt{Z}} \sum_i e^{-\beta (E_i+ \mu Q_i)/2} |E_i, Q_i \rangle_1 \otimes |E_i, -Q_i \rangle_2.
\end{equation}
This is described holographically by a charged or rotating black hole. The holographic correspondence for eternal charged black holes was studied in \cite{Brecher:2004gn,Andrade:2013rra}.  The entanglement structure of these states is similar to the thermofield double, but in the extremal limit $\beta \to 0$, the entanglement becomes more non-local, and the wormhole in the holographic dual becomes infinitely long. The two copies of the CFT remain entangled, however, and there are classes of operators whose two-sided correlation functions remain finite. 

It is interesting to ask how small perturbations of \eqref{ctfd} behave, and how the previous results on the sensitive dependence on initial conditions are modified by the additional scale introduced by the chemical potential. The purpose of this paper is to explore this question holographically, in the simple context of 2+1 gravity, dual to suitable two-dimensional conformal field theories. After completing our work, we realised that this extension of \cite{Shenker:2013pqa}  was previously considered in \cite{Leichenauer:2014nxa}.\footnote{Related work on extending the complexity ideas of  \cite{Susskind:2014rva} to charged black holes appeared in \cite{Halyo:2015fpa}, and the recent work on complexity and action covers both charged and uncharged examples \cite{Brown:2015bva,Brown:2015lvg}.} (Related recent work is \cite{Sircar:2016old,Roberts:2016wdl}.)  There is some overlap with our work; differences are that that paper focuses on the mutual information, and higher-dimensional black holes, while we will focus on correlation functions in 2+1 dimensional black holes. 

We find that the growth of the effect is still  controlled by the temperature; for the case with rotation, the parameter controlling the strength of the shock is 
\begin{equation}
\alpha = \frac{\Delta r_+}{2\kappa l^2}e^{\kappa t_w} = \frac{r_+^2}{(r_+^2 - r_-^2)^2}\left(\frac{E}{4M}(r_+^2 + r_-^2) - \frac{L}{2J}r_-^2\right)\exp\left(\frac{r_+^2 - r_-^2}{l^2r_+}t_w\right), 
\end{equation}
where $\kappa$ is the surface gravity of the black hole and $E$, $L$ are the energy and angular momentum carried by the shock, which modifies the geometry by shifting the outer horizon radius by an amount $\Delta r_+$. This is consistent with the results of \cite{Leichenauer:2014nxa}. The Lyapunov exponent characteristic of quantum chaos is thus still $\lambda_L = \kappa$, as in the simple thermal systems. The prefactor is also controlled by the surface gravity, so the dynamics is not directly sensitive to the additional scale associated with the angular momentum. The same slowing down of time evolution controlled entirely by the temperature is seen in correlation functions on unperturbed charged black holes \cite{Brecher:2004gn,Andrade:2013rra}. 

\section{Review of the uncharged, non-rotating case}

\begin{figure}[tb]
\centering
\includegraphics[width=0.4\textwidth]{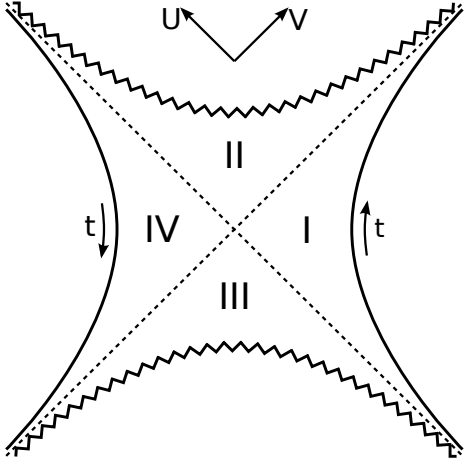}
\caption{Regions I to IV in Kruskal coordinates.}
\label{regions}
\end{figure}

We first review the original work of \cite{Shenker:2013pqa} on the uncharged case. They considered a spherically symmetric perturbation of an uncharged, non-rotating black hole. For simplicity, they considered the non-rotating BTZ solution in 2+1 dimensions, 
\begin{equation}
ds^2 = -f(r)dt^2 + \frac{dr^2}{f(r)} + r^2d\phi^2
\end{equation}
where 
\begin{equation}
f(r) = \frac{r^2 - r_+^2}{l^2} = \frac{r^2}{l^2} - M.
\end{equation}
The horizon radius is $r_+$ (this was $R$ in \cite{Shenker:2013pqa}), $l$ is the AdS scale, and  $M$ is the black hole mass. To understand the matching across the shell near the horizon, we also need to use Kruskal coordinates, 
\begin{align}
U &= -e^{-\kappa u},\\
V &= e^{\kappa v}
\end{align}
where $\kappa = r_+/ l^2$ is the surface gravity, and $u, v = t \mp r_*$, with the tortoise coordinate 
\begin{equation}
r_* = -\int_r^\infty\frac{dr'}{f(r')} = \frac{l^2}{2r_+}\log\left(\frac{r - r_+}{r + r_+}\right). 
\end{equation}
This gives 
\begin{equation}
UV = - \frac{r - r_+}{r + r_+},
\end{equation}
and the manifestly non-singular form of the metric 
\begin{equation}
ds^2 = \frac{-4l^2dUdV + r_+^2(1 - UV)^2d\phi^2}{(1 + UV)^2}.
\label{Kruskal metric}
\end{equation}
This defines the relation of the ordinary BTZ coordinates to Kruskal coordinates in region I of figure \ref{regions}. There are similar relations in the other regions. 

We add energy to the system on the left boundary at some early time, i.e. at a large value, $t_w$, of the $t$ coordinate. For simplicity, it is assumed that the perturbation is spherically symmetric, while the asymptotic energy of the perturbation, $E$, is assumed to be small compared with $M$. Formally, we take a limit $E/ M \rightarrow 0$, $t_w \rightarrow \infty$ with $Ee^{\kappa t_w}/M$ fixed. 

\begin{figure}[tb]
\centering
\includegraphics[width=0.5\textwidth]{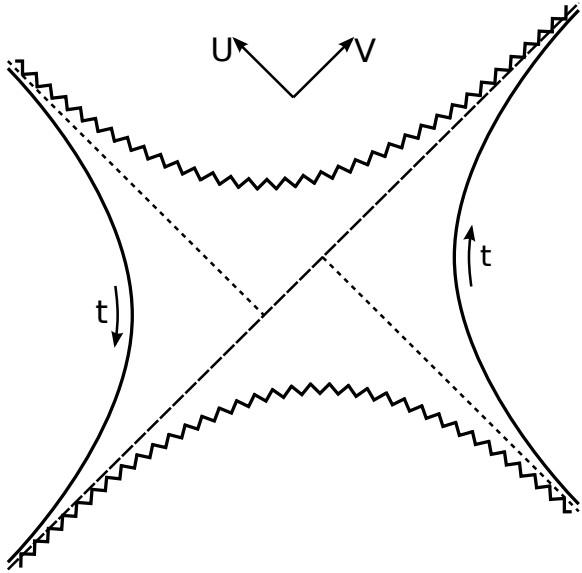}
\caption{Kruskal coordinates and the perturbed BTZ solution.}
\label{perturbed}
\end{figure}

In this limit, the perturbation approximately follows null geodesics, so the perturbed geometry is obtained by gluing a BTZ solution with mass $M$ (to the past/right of the perturbation) to one with mass $M + E$ (to the future/left of the perturbation) across the null surface $U_w = e^{-\kappa t_w}$, which meets the left boundary at $t=t_w$. To the right of the shock we have coordinates $U$, $V$ and $\phi$, with parameter $M$ or $r_+$. To the left, we have coordinates $\tilde{U}$, $\tilde{V}$ and $\phi$ and parameter $\tilde{M} = M + E$ or $\tilde{r}_+ =\sqrt{\frac{M + E}{M}}r_+$. The relationship between the two coordinate systems on the shock is fixed by imposing two conditions:
\begin{enumerate}
\item The time coordinate $t$ is required to be continuous at the boundary, i.e. at $r = \infty$. This fixes a relative boost ambiguity.
\item The size of the $S^1$ must be continuous across the shock. 
\end{enumerate}
The first of these conditions means that, to the left of the shock, $\tilde{U}_w = e^{-\tilde{\kappa} t_w }$, where $\tilde{\kappa} = \tilde{r}_+ / l^2$. In the limit we get $\tilde{U}_w = U_w(1 + t_w\Delta\kappa)$ where $t_w\Delta\kappa$ is small. The second condition then gives 
\begin{equation}
\tilde{V} = V + \alpha,
\label{V has a step}
\end{equation}
where
\begin{equation}
\alpha = \frac{\Delta r_+}{2 \kappa l^2} e^{\kappa t_w} = \frac{E}{4M}e^{r_+t_w / l^2}.
\label{step size}
\end{equation}
This is illustrated in the diagram of figure \ref{perturbed}.

Note that the positivity of $\alpha$, the step change in the $V$ coordinate, is simply related to the second law of thermodynamics for the entropy of the black hole. We can make $\alpha$ as large as desired by pushing the perturbation further back in time, i.e. by increasing $t_w$.

As we will see later, the general form of the perturbation $\alpha$ will be essentially the same in the other cases we consider; the essential ingredients are just the structure of the Kruskal coordinates in terms of the tortoise coordinate and the matching conditions. 


As BTZ is locally AdS$_3$, the length of geodesics is conveniently calculated by using the embedding coordinates in a flat $2 + 2$ dimensional spacetime, in which the length of geodesics between points $p$ and $p'$ is given by 
\begin{equation}
\cosh\frac{d}{l} = \frac{1}{l^2}\left(T_1T'_1 + T_2T'_2 - X_1X'_1 - X_2X'_2\right),
\label{distance formula}
\end{equation}
These coordinates are related to the Kruskal and BTZ coordinates by 
\begin{align}
T_1 &= l\frac{V + U}{1 + UV} = \frac{l}{r_+}\sqrt{r^2 - r_+^2}\sinh\frac{r_+t}{l^2},\\
T_2 &= l\frac{1 - UV}{1 + UV}\cosh\frac{r_+\phi}{l} = \frac{lr}{r_+}\cosh\frac{r_+\phi}{l},\\
X_1 &= l\frac{V - U}{1 + UV} = \frac{l}{r_+}\sqrt{r^2 - r_+^2}\cosh\frac{r_+t}{l^2},\\
X_2 &= l\frac{1 - UV}{1 + UV}\sinh\frac{r_+\phi}{l} = \frac{lr}{r_+}\sinh\frac{r_+\phi}{l},
\end{align}
in region I. We will use a similar method later for rotating BTZ. 

\begin{figure}[htb]
\centering
\includegraphics[width=0.5\textwidth]{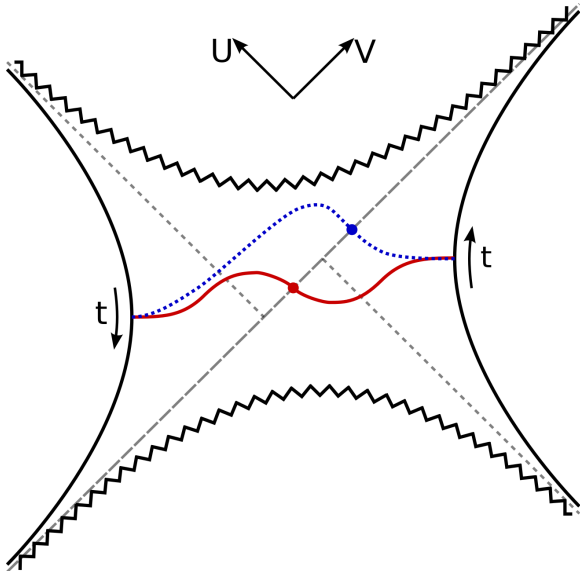}
\caption{Geodesics through the perturbed wormhole are found by gluing geodesics from each side at a general location on the shock and extremizing the total length over this location. The blue, dashed line shows two geodesics glued at an arbitrary location. The red, solid line, passing through the centre of the conformal diagram, extremizes the total length and is therefore the geodesic required.}
\label{gluing geodesics}
\end{figure}

Geodesics between two points on opposite boundaries must necessarily cross the shock. To calculate the geodesic distance between such points, we
\begin{enumerate}
\item Calculate the geodesic distances between a general location, $U = 0$, $V = V_{\text{shock}}$, on the shock and each of the two boundary points.
\item Extremize the sum over $V_{\text{shock}}$.
\end{enumerate}
This is illustrated in figure \ref{gluing geodesics}.
We use the coordinates to the right of the shock to label the point on the shock. If the two boundary points are both at $t = 0$ and at equal angular coordinate $\phi$, we find that the geodesic crosses the shock at the centre of the conformal diagram at $V_{\text{shock}} = -\alpha / 2$, as one would expect from symmetry. Regulating the overall divergence in the length of the geodesic by taking the distance between points at some large fixed radius $r$, we obtain
\begin{equation}
\frac{d}{l} = 2\log\frac{2r}{r_+} + 2\log\left(1 + \frac{\alpha}{2}\right).
\end{equation}
The second term gives the increase in the length of the geodesic resulting from the addition of the perturbation. This increase may be made arbitrarily large by increasing $t_w$, i.e. by adding the perturbation further back in the past. 

As mentioned in the introduction, we can use the geodesic length to obtain an approximation to the two-point correlation function of operators inserted on the two boundaries of the black hole, 
\begin{equation}
\langle W | V_L V_R | W \rangle \sim e^{-\Delta d/l}  \sim r^{-2\Delta} \left(1 + \frac{\alpha}{2}\right)^{-2\Delta},
\end{equation}
where $|W \rangle$ is the state obtained by acting on the thermofield double state with the perturbation $W(t_w)$ on the Hilbert space $\mathcal H_1$. Thinking of the thermofield double state as prepared by a path integral on the Euclidean circle, this correlation function can be interpreted as an out of time order correlation function
\begin{equation}
Z^{-1} \mbox{tr}(e^{-\beta H/2} V W(t_w) e^{-\beta H/2} V W(t_w) ),  
\end{equation}
The exponential growth of $\alpha$ with $t_w$ corresponds to an exponential decay of this OTO correlation function, which leads to a growth in the squared commutator \cite{Maldacena:2015waa}
\begin{equation}
 - Z^{-1} \mbox{tr}e^{-\beta H/2} [W(t_w),V] e^{-\beta H/2} [W(t_w),V] ). 
\end{equation}
The time scale at which $\alpha$ becomes of order one, $t_s = \frac{\beta}{2\pi} \ln \frac{M}{E}$, is recognised as the scrambling time.  On this same time scale, the entanglement between the two copies of the CFT in the thermofield double state, which was initially between approximately local degrees of freedom in the two copies, has become delocalised.  

\section{Perturbing the rotating BTZ solution}

The simplest extension of this calculation to consider is the rotating BTZ solution, as the geometry is still locally AdS, so geodesic calculations will be simple, and a good deal of progress can be made analytically. This introduces an additional length scale associated with the rotation, and the interesting question is to what extent the physical effects depend on this scale. 

The rotating BTZ metric is 
\begin{equation}
ds^2 = -f^2(r)dt^2 + f^{-2}(r)dr^2 + r^2\left[N^\phi(r)dt + d\phi\right]^2
\end{equation}
where
\begin{equation}
f^2(r) = -M + \left(\frac{r}{l}\right)^2 + \frac{J^2}{4r^2},
\end{equation}
and we adopt co-rotating coordinates, since we are interested in the behaviour near the horizon, so 
\begin{equation}
N^\phi(r) = \frac{J}{2}\frac{r^2 - r_+^2}{r^2r_+^2}.
\end{equation}
The horizon radii $r_+$ and $r_-$ are the solutions to $f^2(r) = 0$, 
\begin{equation}
r_\pm^2 = \frac{1}{2}\left(Ml^2 \pm \sqrt{(Ml^2)^2 - J^2l^2}\right).
\end{equation}
We will find it useful to express the metric entirely in terms of $r_+$ and $r_-$, rather than $M$ and $J$, which are 
\begin{equation}
\begin{split}
M &= \frac{r_+^2 + r_-^2}{l^2},\\
J &= \frac{2r_+r_-}{l}.
\end{split}
\label{M and J}
\end{equation}
We will assume without loss of generality that $J$ is positive. The metric functions in terms of $r_\pm$ are 
\begin{equation}
N^\phi(r) = \frac{r_-}{r_+}\,\frac{r^2 - r_+^2}{lr^2}
\end{equation}
and
\begin{equation}
f^2(r) = \frac{(r^2 - r_+^2)(r^2 - r_-^2)}{l^2r^2}.
\end{equation}

We introduce Kruskal coordinates by writing as before 
\begin{align}
U &= -e^{-\kappa u},\\
V &= e^{\kappa v},
\end{align}
where $u, v = t \mp r_*$ and the tortoise coordinate is 
\begin{equation}
r^* = \frac{1}{2\kappa}\log\frac{\sqrt{r^2 - r_-^2} - \sqrt{r_+^2 - r_-^2}}{\sqrt{r^2 - r_-^2} + \sqrt{r_+^2 - r_-^2}}
\end{equation}
where $\kappa$ is given by
\begin{equation}
\kappa = \frac{r_+^2 - r_-^2}{l^2r_+}.
\end{equation}
This gives the metric 
\begin{equation}
ds^2 = \frac{-4l^2dUdV - 4lr_-(UdV - VdU)d\phi + \left[(1 - UV)^2r_+^2 + 4UVr_-^2\right]d\phi^2}{(1 + UV)^2}.
\end{equation}
%


\subsection{Adding the perturbation}

We consider a spherically symmetric shell which meets the left boundary at some time $t_w$. For finite $t_w$, the trajectory of this shell in the $U,V$ plane will depend on the angular momentum it carries, but as we take the limit of large $t_w$, we apply a large boost in the $U,V$ plane, and the trajectory becomes approximately lightlike, along a line of constant $U$, as in the non-rotating case.  The matching problem is then very similar to the one in the non-rotating case. We glue two copies of the rotating BTZ spacetime together along a shock at $U = e^{-\kappa t_w}$. To the right of the shock, the black hole has mass $M$, angular momentum $J$ and coordinates $U$, $V$ and $\phi$, while to the left of the shock, we have mass $\tilde{M} = M + E$, angular momentum $\tilde{J} = J + L$ and coordinates $\tilde{U}$, $\tilde{V}$ and $\phi$. We impose continuity of $t$ at the boundary and $r$ across the shock as for the non-rotating case. The result is a jump in $V$, 
\begin{equation}
\tilde{V} = V + \alpha,
\end{equation}
where
\begin{equation}
\alpha = \frac{\Delta r_+}{2\kappa l^2}e^{\kappa t_w}
\end{equation}
exactly as in the non-rotating case. In terms of $M$, $J$, $E$ and $L$
\begin{equation}
\alpha = \frac{r_+^2}{(r_+^2 - r_-^2)^2}\left(\frac{E}{4M}(r_+^2 + r_-^2) - \frac{L}{2J}r_-^2\right)\exp\left(\frac{r_+^2 - r_-^2}{l^2r_+}t_w\right).
\end{equation}

Since the rotating black holes have a throat which grows infinitely long in the extremal limit, one might have thought that for near-extremal black holes it would be possible to add a shock that took one away from extremality, increasing the size of the black hole while lowering the length of the wormhole. However, we find that so long as the second law of thermodynamics is obeyed so $\Delta r_+ >0$, the jump $\alpha >0$. We will now see that this leads to a longer wormhole. 


\subsection{Geodesic lengths}

We will calculate the length of the geodesics in embedding coordinates, as in the non-rotating case. For our co-rotating coordinates, the relation to embedding coordinates is
\begin{align}
T_1 &= \pm\sqrt{\pm B(r)}\sinh\tilde{t}(t, \phi),\\
T_2 &= \sqrt{A(r)}\cosh\tilde{\phi}(t, \phi),\\
X_1 &= \pm\sqrt{\pm B(r)}\cosh\tilde{t}(t, \phi),\\
X_2 &= \sqrt{A(r)}\sinh\tilde{\phi}(t, \phi),
\end{align}
where
\begin{align}
A(r) &= l^2\frac{r^2 - r_-^2}{r_+^2 - r_-^2},\\
B(r) &= l^2\frac{r^2 - r_+^2}{r_+^2 - r_-^2}
\end{align}
and
\begin{align}
\tilde{\phi} &= \frac{r_+\phi}{l}\\
\tilde{t} & = \kappa t - \frac{r_-}{l}\phi.
\end{align}
Here the first $\pm$ in the formulae is positive for regions I and II and negative for regions III and IV, while the second is positive for regions I and IV and negative for regions II and III. The transformation from the Kruskal coordinates to the embedding coordinates is 
\begin{equation}
\begin{split}
T_1 &= l\frac{V + U}{1 + UV}\cosh\frac{r_-\phi}{l} - l\frac{V - U}{1 + UV}\sinh\frac{r_-\phi}{l},\\
T_2 &= l\frac{1 - UV}{1 + UV}\cosh\frac{r_+\phi}{l},\\
X_1 &= l\frac{V - U}{1 + UV}\cosh\frac{r_-\phi}{l} - l\frac{V + U}{1 + UV}\sinh\frac{r_-\phi}{l},\\
X_2 &= l\frac{1 - UV}{1 + UV}\sinh\frac{r_+\phi}{l}.
\end{split}
\label{embedding}
\end{equation}
These hold in each of the four regions.

We consider first a geodesic from a point at $t=0, \phi=0$ on one boundary to a point at $t=0, \phi =0$ on the other boundary. The main complication relative to the discussion in \cite{Shenker:2013pqa} is that the geodesic may not meet the shock at $\phi =0$. We must join geodesics from the two boundary points at a general point on the shock and then extremize the geodesic length with respect to both the $V$ and $\phi$ coordinates of the meeting point.

To the left of the shock, we need the distance from $(t, r, \phi) = (0, r, 0)$ (in region IV) to $(U', V', \phi') = (0, V + \alpha, \phi)$. The embedding coordinates of the first point are
\begin{align}
T_1 &= 0,\\
T_2 &= l\sqrt{\frac{r^2 - r_-^2}{r_+^2 - r_-^2}},\\
X_1 &= -l\sqrt{\frac{r^2 - r_+^2}{r_+^2 - r_-^2}},\\
X_2 &= 0.
\end{align}
while for the second point we get
\begin{align}
T'_1 &=  l(V + \alpha)\cosh\frac{r_-\phi}{l} - l(V + \alpha)\sinh\frac{r_-\phi}{l} = l(V + \alpha)e^{-r_-\phi/l},\\
T'_2 &= l\cosh\frac{r_+\phi}{l},\\
X'_1 &= l(V + \alpha)\cosh\frac{r_-\phi}{l} - l(V + \alpha)\sinh\frac{r_-\phi}{l} = l(V + \alpha)e^{-r_-\phi/l},\\
X'_2 &= l\sinh\frac{r_+\phi}{l}
\end{align}
If we let $d_1$ be the length of the geodesic to the left of the shock, then
\begin{align}
\cosh\frac{d_1}{l} &= \frac{1}{l^2}(T_2T'_2 - X_1X'_1)\\
&= \sqrt{\frac{r^2 - r_-^2}{r_+^2 - r_-^2}}\cosh\frac{r_+\phi}{l} + (V + \alpha)\sqrt{\frac{r^2 - r_+^2}{r_+^2 - r_-^2}}e^{-r_-\phi / l}\\
&\simeq \frac{r}{\sqrt{r_+^2 - r_-^2}}\left(\cosh\frac{r_+\phi}{l} + (V + \alpha)e^{-r_-\phi / l}\right).
\end{align}
For the geodesic to the right of the shock, the calculation proceeds as above, but with the sign of $X_1$ reversed for the boundary point and $V + \alpha$ replaced by $V$ at the shock. Hence
\begin{equation}
\cosh\frac{d_2}{l} \simeq \frac{r}{\sqrt{r_+^2 - r_-^2}}\left(\cosh\frac{r_+\phi}{l} - Ve^{-r_-\phi / l}\right).
\label{d2}
\end{equation}
To find the value of $V$ that extremizes $d = d_1 + d_2$, we differentiate to get
\begin{align}
\frac{1}{l}\sinh\left(\frac{d_1}{l}\right)\frac{\partial d_1}{\partial V} &= \frac{r}{\sqrt{r_+^2 - r_-^2}}e^{-r_-\phi / l},\\
\frac{1}{l}\sinh\left(\frac{d_2}{l}\right)\frac{\partial d_2}{\partial V} &= -\frac{r}{\sqrt{r_+^2 - r_-^2}}e^{-r_-\phi / l}
\end{align}
so that
\begin{equation}
\frac{\partial d}{\partial V} = \frac{lr}{\sqrt{r_+^2 - r_-^2}}e^{-r_-\phi / l}\left(\frac{1}{\sinh\frac{d_1}{l}} - \frac{1}{\sinh\frac{d_2}{l}}\right).
\end{equation}
This vanishes if $d_1 = d_2$, which gives $V = -\alpha / 2$, as we might again have expected from symmetry.  Equation (\ref{d2}) now gives us
\begin{equation}
\frac{d}{2l} = \log\frac{2r}{\sqrt{r_+^2 - r_-^2}} + \log\left(\cosh\frac{r_+\phi}{l} + \frac{\alpha}{2}e^{-r_-\phi / l}\right),
\label{lengthening}
\end{equation}
where we have used $\cosh^{-1}x \simeq \pm\log(2x)$ for large $x$. Note that since $\alpha > 0$, the perturbation must increase the length of the geodesic, as we said earlier. 

Extremizing (\ref{lengthening}) with respect to $\phi$ gives us
\begin{equation}
r_+\sinh\frac{r_+\phi}{l} = \frac{\alpha r_-}{2}e^{-r_-\phi / l}.
\label{joining phi}
\end{equation}
We define $\phi^*$ to be the value of $\phi$ satisfying this equation and we let $p(\alpha)$ be the contribution of the perturbation to the geodesic length $d/l$, i.e.
\begin{equation}
p(\alpha) = 2\log\left(\cosh\frac{r_+\phi^*}{l} + \frac{\alpha}{2}e^{-r_-\phi^* / l}\right),
\end{equation}
so that
\begin{equation} \label{chgeod}
\frac{d}{l} = 2\log\frac{2r}{\sqrt{r_+^2 - r_-^2}} + p(\alpha).
\end{equation}

\begin{figure}[tb]
\centering
\includegraphics[width=0.7\textwidth]{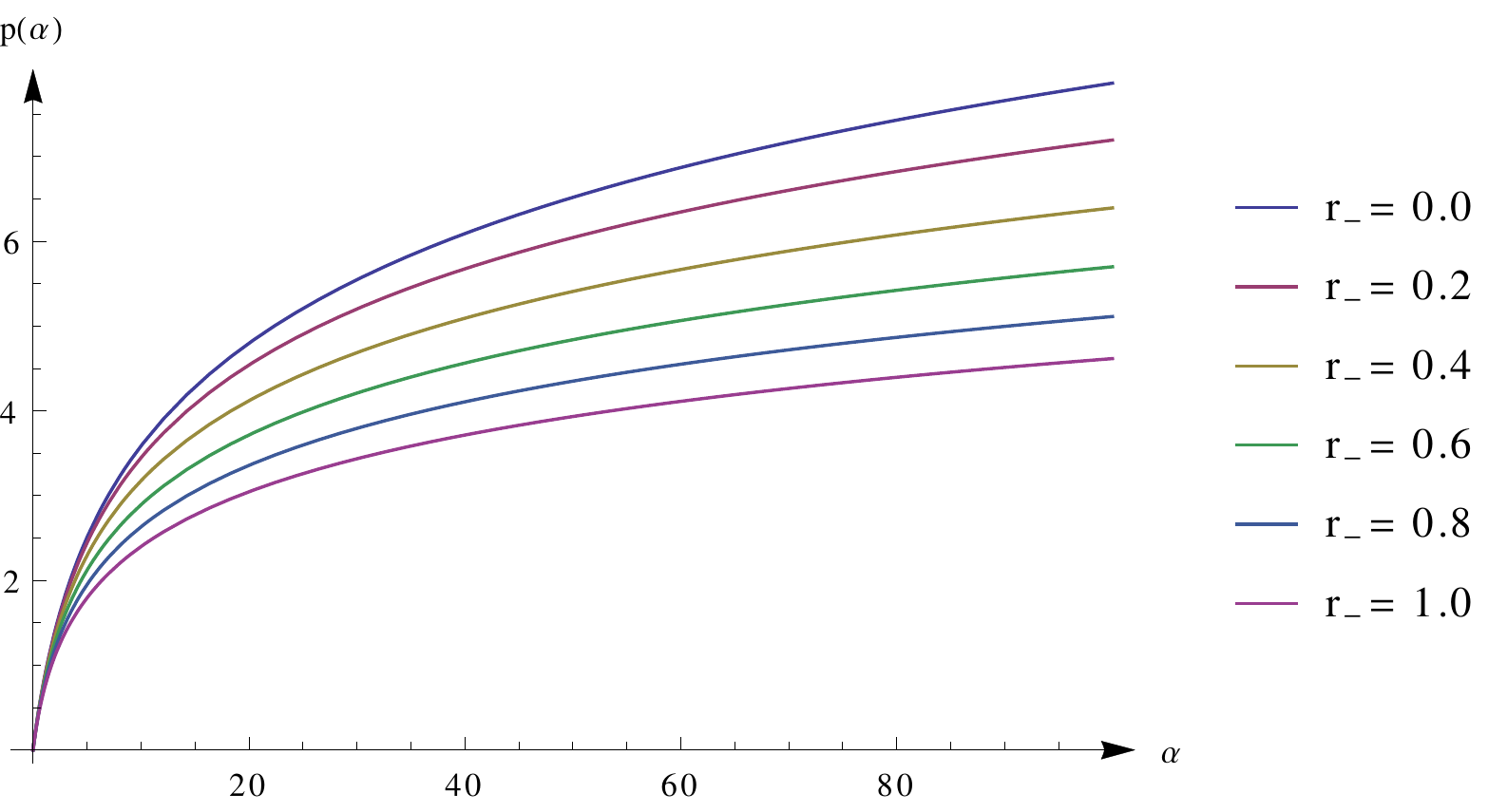}
\caption{Increase in the length of the geodesic, as a function of the size of the jump in $V$ coordinate at the shock. Here $r_+ = 1$, so the plot shows graphs for the non-rotating and extremal black holes and four intermediate cases.}
\label{p}
\end{figure}

Unfortunately, it appears that we cannot solve (\ref{joining phi}) analytically, except in the special cases of non-rotating and extremal black holes. In the first case, we saw earlier that $\phi^* = 0$ and
\begin{equation}
p(\alpha) = 2\log\left(1 + \frac{\alpha}{2}\right), 
\end{equation}
while for extremal black holes when $r_+ = r_-$ we get
\begin{equation}
 \phi^* = \frac{l}{2r_+}\log\left(1 + \alpha\right)
\end{equation}
and
\begin{equation}
p(\alpha) = \log(1 + \alpha).
\end{equation}
In the general case, it is straightforward to show that both $\phi^*$ and $p(\alpha)$ (and hence the geodesic length) increase with $\alpha$. Given the expressions for $p(\alpha)$ for the two special cases, one would expect similar logarithmic increases in $p(\alpha)$ with respect to $\alpha$ in the general case. The results of numerical calculations, displayed in figure \ref{p},
would appear to confirm this.

\begin{figure}[tb]
\centering
\includegraphics[width=0.7\textwidth]{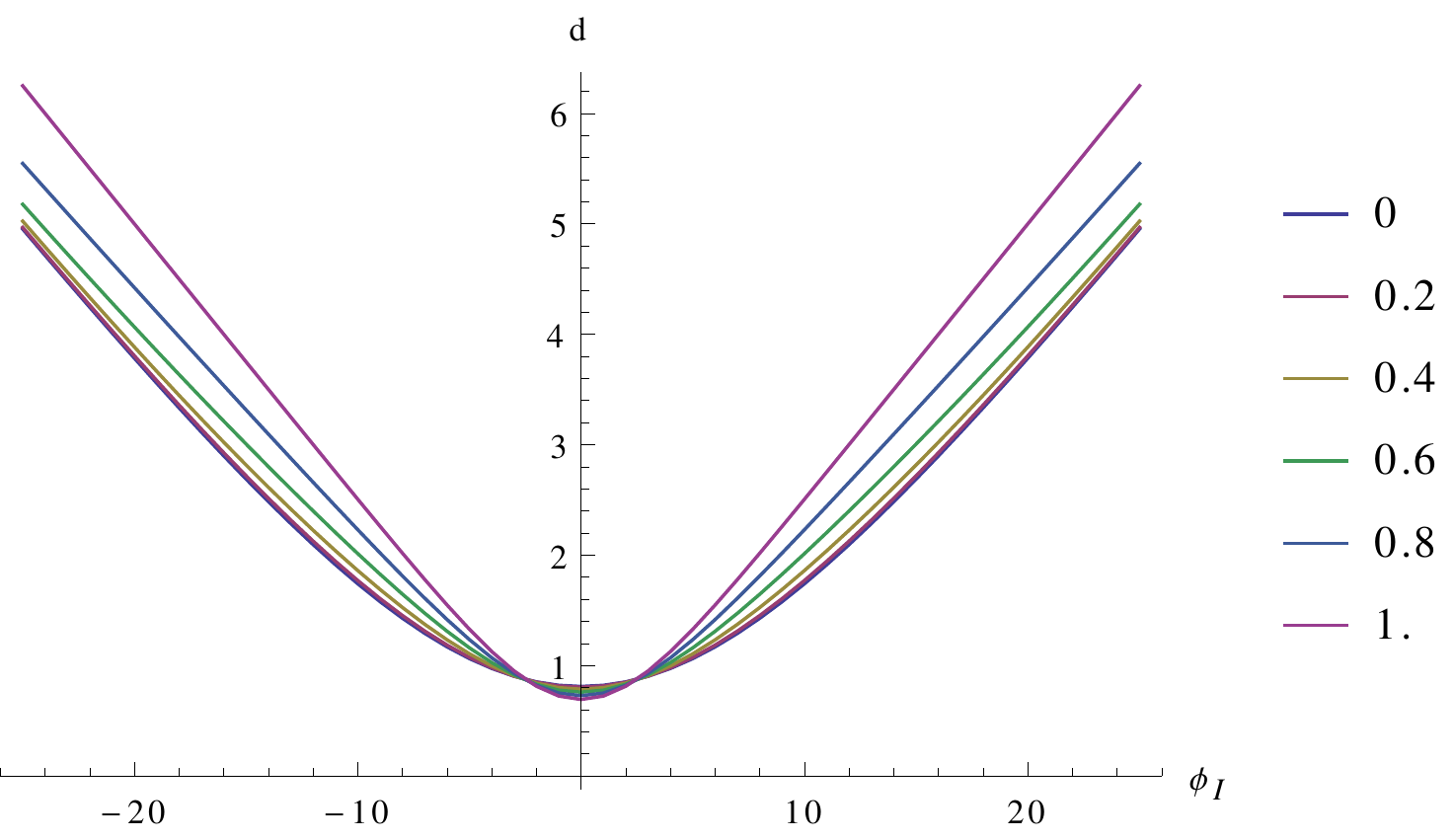}
\caption{Overall geodesic distance plotted against the value of $\phi$ on the right hand boundary, for a range of different $r_-$. Again, $r_+ = 1$, while $\phi$ is set to zero on the left hand boundary.}
\label{dVsPhi}
\end{figure}

Given the non-trivial behaviour of the angular coordinate for these geodesics, there is the concern that it might be possible to find a shorter geodesic between the boundary points, by allowing $\phi$ to go from zero on one boundary to $\phi = 2\pi$ on the other. Applying the numerical calculations to general values of $\phi$ on the boundaries is straightforward, resulting in figure \ref{dVsPhi}.
The monotonic increase in geodesic length with the difference in angular coordinate confirms that the shortest geodesic between matching boundary points is that calculated between matching values of $\phi$, not values differing by some multiple of $2\pi$. 

As in the non-rotating case, this increase in the length of the geodesics can be interpreted as a decrease in the correlation functions of operators in the state $|W \rangle$ created by acting with the perturbation $W(t_w)$. In the rotating black hole, the initial value of the correlators before the perturbation is smaller, as the factor of $\sqrt{r_+^2 -r_-^2}$ in \eqref{chgeod} increases the length of the geodesics, but the dynamical evolution is as in the non-rotating case, and the change in the length of the geodesics becomes appreciable when $\alpha$ is of order one, at the scrambling time $t_s = \kappa^{-1} \ln \kappa/\Delta r_+$. As in the non-rotating case, this scales as the ratio of the energy of the black hole to the energy of the perturbation. If we take the extremal limit, $\Delta r_+$ could be small compared to $r_+$ but large compared to $\kappa$, but for this to change the scaling form of $t_s$ we would need to go to temperatures $T$ of order the energy of the perturbation.   

We can also consider the implications of the geodesic calculation for the entanglement entropy (as in \cite{Leichenauer:2014nxa}), which is also similar to the non-rotating case. Consider the mutual information of two matching regions, one on each boundary, with arc length $\phi$ and centred on the same angular coordinate. Firstly, the entanglement entropy of one of the regions is, assuming $\phi < \pi$,
\[
S_A = \frac{l}{4G_N}\left(2\log\frac{2r}{\sqrt{r_+^2 - r_-^2}} + \log\sinh\frac{(r_+ + r_-)\phi}{2l} + \log\sinh\frac{(r_+ - r_-)\phi}{2l}\right).
\]
Meanwhile, the entanglement entropy of $A \cup B$ is the smallest of
\begin{align}
S_{A \cup B}^{(1)} &= S_A + S_B,\\
S_{A \cup B}^{(2)} &= \frac{l}{2G_N}\left(2\log\frac{2r}{\sqrt{r_+^2 - r_-^2}} + p(\alpha)\right).
\end{align}
Now
\begin{equation}
S_{A \cup B}^{(1)} - S_{A \cup B}^{(2)} = \frac{l}{2G_N}\left(\log\sinh\frac{(r_+ + r_-)\phi}{2l} + \log\sinh\frac{(r_+ - r_-)\phi}{2l} - p(\alpha)\right)
\label{mutual information}
\end{equation}
and if this is positive, then it gives the mutual information, $I(A; B)$. Otherwise, the mutual information is zero and there is no entanglement between the two regions. Near extremality, we need to have large regions to have non-zero mutual information. But our interest here is in the effect of the perturbation, and again the effect becomes significant, decreasing the local entanglement, just when $\alpha$ becomes of order one.  Local entanglement is therefore reduced by the perturbation at a rate controlled by the scrambling time. 



\section{Perturbing the charged BTZ solution}

The calculation for rotating BTZ is interesting, but as the solution is still locally AdS$_3$, this is a rather special case. We would like to extend the above calculation to further examples. As we will discuss in the next section, considering the correlators for black holes in higher dimensions (charged or uncharged) is challenging. Therefore, we consider here the calculation for a charged black hole in $2+1$ dimensions. We consider Einstein-Hilbert gravity coupled to an ordinary Maxwell field. (It is perhaps more common to consider a Chern-Simons gauge field in this context, but then the solution would remain locally AdS.) 

The metric is 
\begin{equation}
ds^2 = -f(r)dt^2 + \frac{dr^2}{f(r)} + r^2d\phi^2
\end{equation}
where
\begin{equation}
f(r) = \frac{r^2}{l^2} - M - \frac{Q^2}{2}\log\frac{r}{l}.
\end{equation}
This is supported by a gauge field 
\begin{equation}
A = Q \log \frac{r}{r_+} dt. 
\end{equation}

We can introduce Kruskal coordinates where 
\begin{align}
U &= -e^{-\kappa u},\\
V &= e^{\kappa v},
\end{align}
where $u, v = t \mp r_*$ with a tortoise coordinate 
\begin{equation} \label{tcharged}
r_* = - \int_r^\infty \frac{dr}{f(r)}, 
\end{equation}
and $\kappa = f'(r_+) / 2$ is the surface gravity. The metric in these coordinates is 
\begin{equation} \label{kcharged}
ds^2 = \frac{f(r)}{\kappa^2UV}dUdV + r^2d\phi^2,
\end{equation}
where $r$ is a function of $U$ and $V$. In this case one cannot evaluate the integral in \eqref{tcharged} for the tortoise coordinate, so we cannot give a simple expression for $r$ in terms of $U$ and $V$. Near the horizon $r=r_+$, 
\begin{equation} 
\lim_{r \to r_+} r_*=  \frac{1}{2\kappa} \ln \left( \frac{r-r_+}{r_+} \right) + \frac{1}{2\kappa} \ln C
\end{equation}
for some finite constant $C$. This gives $UV \approx -C\frac{(r-r_+)}{r_+}$, so the metric \eqref{kcharged} is regular there. The constant $C$ can be determined numerically for generic parameter values; in the extremal limit $r_+ \to r_-$, it diverges as $C \sim 1/(r_+ - r_-) \sim 1/\kappa$, as in the rotating case.


We consider perturbing this solution by throwing in a charged spherically symmetric shell from the left boundary at some early time $t_w$. The shell will then approximately follow the null trajectory $U = e^{-\kappa t_w}$. The  step change in the $V$ coordinate in the shock is determined by the same matching conditions, which give, as before
\begin{equation}
\tilde{V} = V + \alpha,
\end{equation}
where
\begin{equation}
\alpha = C\frac{\Delta r_+}{r_+} e^{\kappa t_w}.
\end{equation}
Here the relation between $\Delta r_+$ and the parameters of the shell would need to be determined numerically for finite $\Delta r_+$ --- for small perturbations adding $m$ to the black hole mass $M$ and $q$ to the charge $Q$ we have
\begin{equation}
\Delta r_+ \approx \frac{1}{2\kappa}\left(m + Qq\log\frac{r_+}{l}\right).
\end{equation}
 However, we can see that positivity of the shift $\alpha$ continues to be related to the second law. 

\subsection{Geodesic lengths}
\label{charged no perturbation}

For this case, we cannot find the lengths of geodesics by using the embedding coordinates, so we need to simply solve the geodesic equations numerically. Using the symmetry of the solution we can reduce the problem to an effective one-dimensional problem, for spacelike geodesics 
\begin{equation}
\dot{r}^2 = f(r) \left(1 - \frac{L^2}{r^2} \right) + E^2,
\end{equation}
where $E = f(r) \dot t$ and $L = r^2 \dot \phi$ are the constants of motion. 

In the unperturbed spacetime, we are interested in geodesics in a constant-time slice (at $t=0$), so we take $E=0$.  These geodesics can have turning points at $r = r_-$, $r = r_+$ or $r = \left|L\right|$.  For $\left|L\right| > r_+$ we obtain geodesics that return to the boundary from which they started. These will be used in calculations of mutual information. Smaller values of $\left|L\right|$ pass through the wormhole. In either case, half the geodesic length is given by
\begin{equation}
\frac{d}{2} = \lambda_{\text{turn}} = \int_\infty^{r_{\text{turn}}} \frac{dr}{\dot{r}} = \bigints_{r_{\text{turn}}}^\infty \frac{dr}{\sqrt{\left(1 - \frac{L^2}{r^2}\right)f(r) }},
\end{equation}
where we have assumed that the affine parameter $\lambda$ starts at zero on the boundary and that $\dot{r}$ is negative up to the half way point at $\lambda = \lambda_{\text{turn}}$. This is clearly divergent. To find the convergent part, we calculate the integral up to some large value $R$ and subtract the divergent part, given by $l\log R$. We also need to determine the change in the angular coordinate, 
\begin{equation}
\frac{\Delta\phi}{2} = \int_0^{\lambda_{\text{turn}}} \frac{L}{r^2}\,d\lambda.
\end{equation}

For the perturbed spacetime, we consider the geodesics connecting two points at $t=0, \phi=0$ on the two boundaries. The symmetry implies the minimal geodesic connecting these points will have $L=0$. It will run from the first boundary to some point on the shock with arbitrary $V$ coordinate and then to the second boundary; we need to consider general points on the shock and extremise over the position. These geodesics will then have $E \neq 0$. The turning points are solutions to $f(r) + E^2 = 0$. If $E$ is large enough then there are no solutions, and the geodesic hits the singularity. Alternatively, there will be two (possibly coincident) solutions, with values of $r$ between $r_-$ and $r_+$. 

\begin{figure}[htb]
\centering
\includegraphics[width=0.5\textwidth]{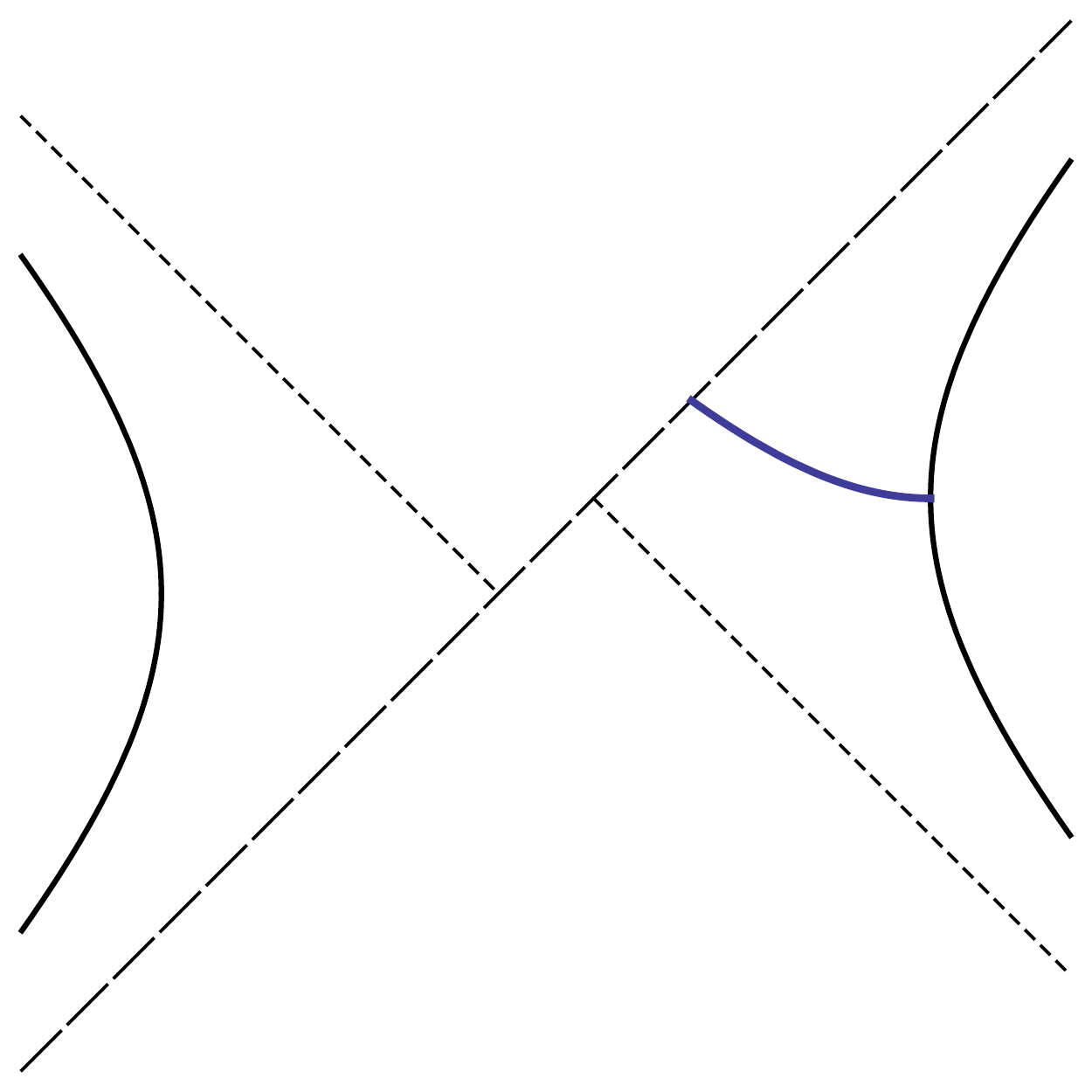}
\caption{Geodesic for $L = 0$, $E = 1$, up to the shock. Note that $r$ remains greater than $r_+$ until the geodesic reaches the shock.}
\label{positive H}
\end{figure}

The simplest case is when $E > 0$. Then $\dot{t} > 0$ and so the geodesic reaches the shock at $r = r_+$ before reaching a turning point. Using 
\begin{equation}
\dot{v} = \frac{E - \sqrt{\left(1 - \frac{L^2}{r^2}\right)f(r) + E^2}}{f(r)}, 
\label{v differential equation}
\end{equation}
given a solution for $r$, we can integrate to obtain $v$ at the intersection with the shock.  The geodesic for $E = 1$, up to the shock, is shown in figure \ref{positive H}.

The more important case will be when $E<0$, so that the geodesic passes through the past horizon at $r = r_+$, and reaches a turning point where $\dot{r}$ becomes positive before reaching the shock. These will give the minimum length geodesics. The geodesic must then be calculated in two halves, before and after the turn. Up to the turn we solve for $r$ and $u$ in terms of $\lambda$ as before, but using
\begin{equation}
\dot{u} = \frac{E + \sqrt{\left(1 - \frac{L^2}{r^2}\right)f(r) + E^2}}{f(r)}
\end{equation}
to calculate $u$ rather than $v$ since $v$ behaves poorly upon crossing the past horizon. We then convert $u$ to $v$ at the turn by adding $2r^*$ using the region III formula for $r^*$.

\begin{figure}[htb]
\centering
\includegraphics[width=0.5\textwidth]{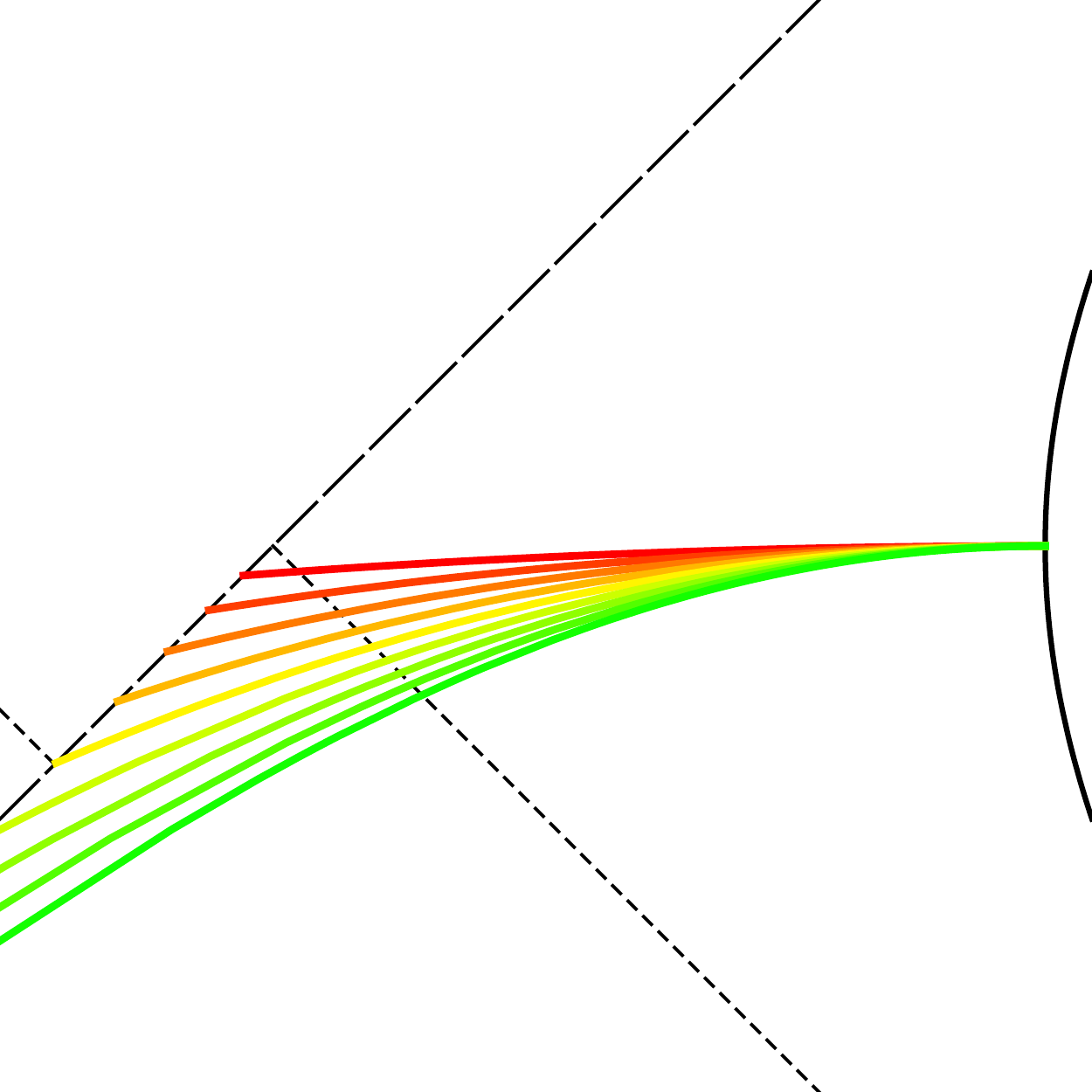}
\caption{Geodesics for $L = 0$ and $E = -0.05, -0.1, \dots, -0.45$. The red, topmost geodesic is for $E = -0.05$, with the hue changing gradually as $-E$ increases. Note that the geodesics all pass through the past horizon and reach a turning point before hitting the shock.}
\label{negative H}
\end{figure}

To handle the second half, we integrate $dr/\dot{r}$ from $r_{\text{turn}}$ to $r_+$ to get $\lambda$ at the shock, and hence the length of the geodesic up to this point. We numerically solve the differential equation for $r$ back from the shock to the turning point and use the result to solve for $v$ using (\ref{v differential equation}). Geodesics for a range of negative values of $E$ are shown in figure \ref{negative H}.

\begin{figure}[htb]
\centering
\includegraphics[width=0.5\textwidth]{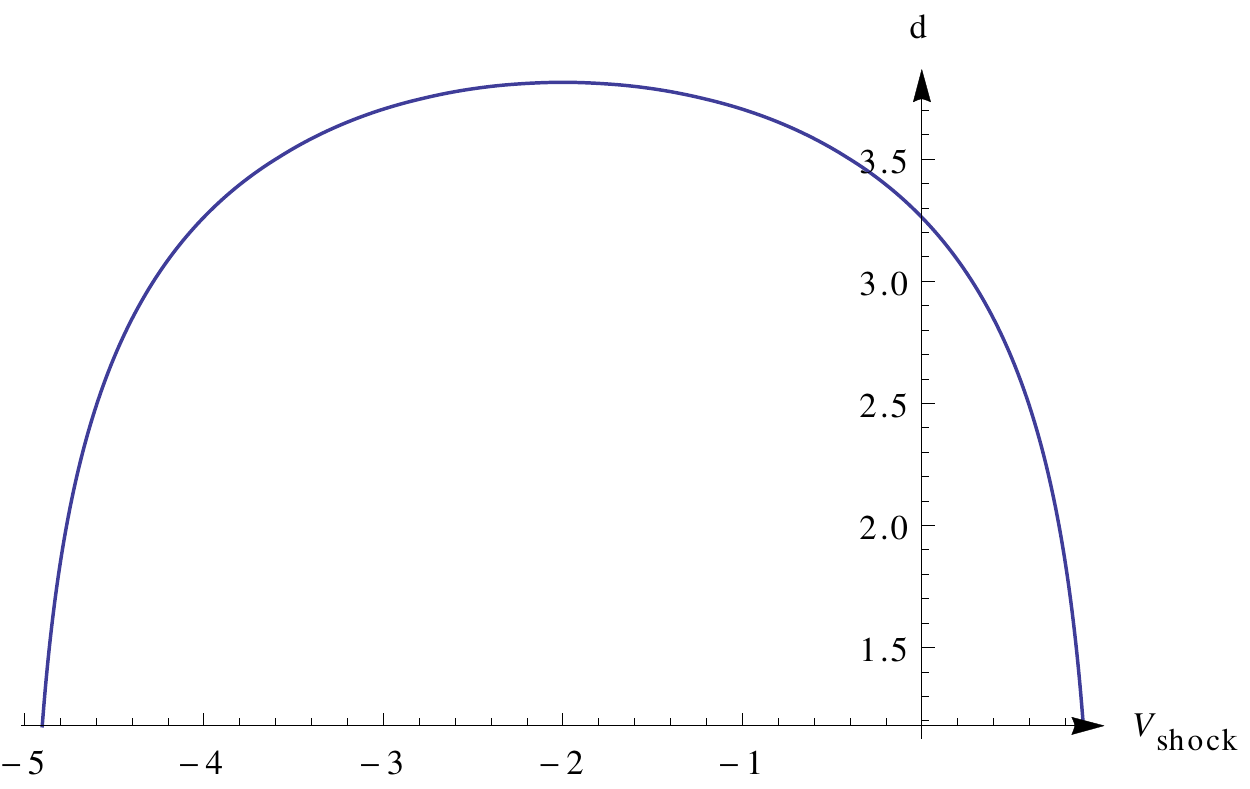}
\caption{The sum of geodesic lengths from $(U, V, \phi) = (-1, 1, 0)$ to $(U, V, \phi) = (0, V_{\text{shock}}, 0)$ to the right of the shock and $(U, V, \phi) = (1, -1, 0$ to $(U, V, \phi) = (0, V_{\text{shock}} + \alpha, 0)$ to the left, plotted against $V_{\text{shock}}$ for $\alpha = 4$. Geodesic length is extremal only at the centre of the perturbed conformal diagram at $V_{\text{shock}} = -\alpha / 2$.}
\label{alpha4}
\end{figure}

\begin{figure}[htb]
\centering
\includegraphics[width=0.5\textwidth]{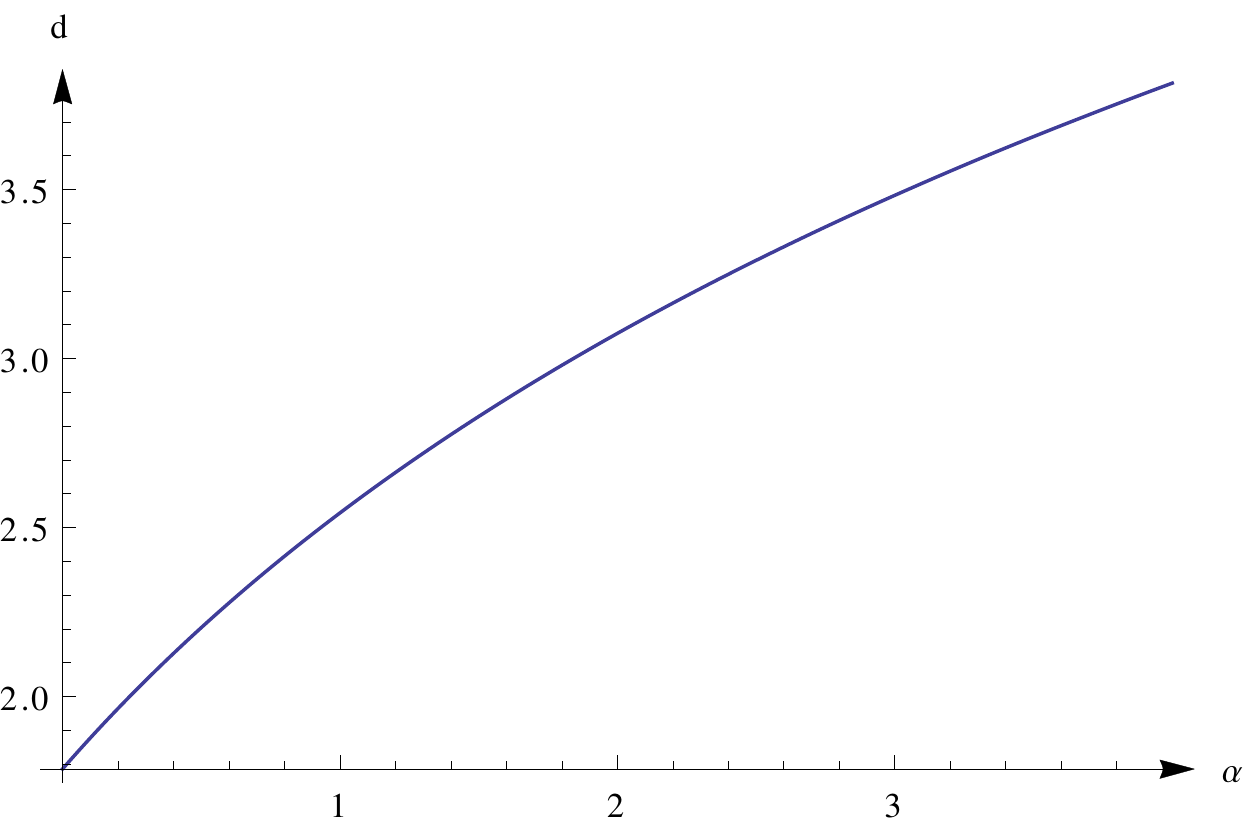}
\caption{Geodesic length across the shock, plotted against the strength of the shock as given by $\alpha$.}
\label{length vs alpha}
\end{figure}

If we now calculate geodesics for a sufficient number of values of $E$ then we can estimate the value of $E$ required to hit any particular point on the shock. This allows us to calculate the length of full geodesics across the shock. If the perturbation gives a step change of $\alpha$ in the $V$ coordinate upon crossing the shock, then for each value of $V_{\text{shock}}$ we sum the length of geodesics from the right boundary to $(U, V) = (0, V_{\text{shock}})$ and from the left boundary to $(U, V) = (0, V_{\text{shock}} + \alpha)$. If we do this for, for example, $\alpha = 4$, then we obtain the results in figure \ref{alpha4}.
The lack of any extrema except for the one expected by symmetry, at $V_{\text{shock}} = -\alpha / 2$, repeated for other values of $\alpha$ indicates that the geodesics joining matching points on the two boundaries cross the shock at the centre of the conformal diagram. This allows us to easily plot the geodesic length against $\alpha$, as in figure \ref{length vs alpha}.

We see that the geodesic length increases monotonically with $\alpha$, becoming significant only for $\alpha$ of order one. Thus, as in the rotating case, the effect of the perturbation on correlation functions and mutual information at $t=0$ is determined by the scrambling time at which $\alpha$ becomes of order one, $t_s \sim \kappa^{-1} \ln r_+/\Delta r_+ \sim \kappa^{-1} \ln N^2$. 

\section{Higher dimensions}

Our investigations, and the original work on the butterfly effect in \cite{Shenker:2013pqa}, have focused on black holes in three bulk dimensions, corresponding to two-dimensional field theories. It would seem useful to extend the discussion to higher dimensions, as in the study of mutual information in \cite{Leichenauer:2014nxa}. However, there is a significant obstacle to doing so for correlation function calculations. In more than three bulk dimensions, the correlation functions in the unperturbed thermofield double state for $t \neq 0$ are not correctly reproduced by considering the real geodesics in the real Lorentzian geometry; one needs to take complexified geodesics into account \cite{Fidkowski:2003nf}. The geodesics in the real Lorentzian geometry become null, corresponding to a singular correlation function, if we consider equal-time correlation functions at some boundary time $t = -t_*$. 

The correlations we have been considering in the perturbed black hole are at $t=0$, but the calculation involves a geodesic on the right which goes from $t=0$ on the boundary to a point on the shock at $V = -\alpha/2$ (and on the left, from $t=0$ on the boundary to a point on the shock at $V = \alpha/2$). If we considered extending this geodesic to the other boundary in the unperturbed geometry, it would meet the other boundary at some $t= -t_0$. Thus, this is just a time-translated version of the geodesic that \cite{Fidkowski:2003nf} concluded was not relevant to the calculation of the correlator on the real sheet. 

\begin{figure}[htb]
\centering
\includegraphics[width=0.5\textwidth]{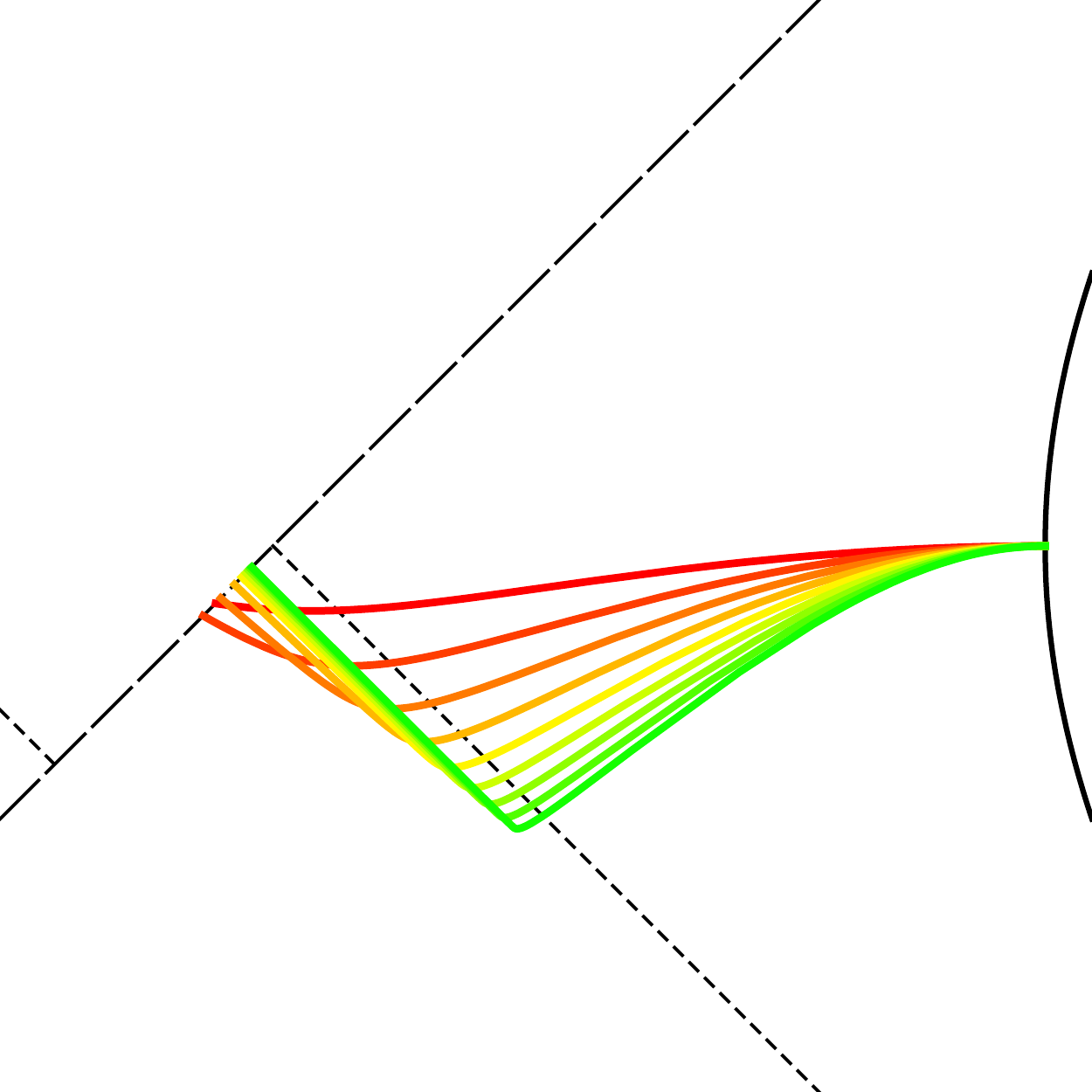}
\caption{Geodesics for $L = 0$ and $E = -0.5, -1, \dots, -4.5$ for the simple 3+1 dimensional black hole. The black hole mass, $M$, and the AdS length, $l$, are both set to 1. The red geodesic is for $E = -0.5$, with the hue changing gradually as $-E$ increases. As $-E$ increases, the intersection with the shock moves away from $V = 0$, reaches a critical point and then moving back towards, but does not reach, $V = 0$.}
\label{4d negative H}
\end{figure}

This is signalled by the fact that when we consider the geodesic from the boundary to the shock as a function of $\alpha$, there is a critical value of $\alpha$ beyond which there is no longer a spacelike geodesic which connects $t=0$ on the boundary to $V = -\alpha/2$ on the shock, as shown in figure \ref{4d negative H}. This critical value of $\alpha$ should correspond to the critical time $t_*$ in \cite{Fidkowski:2003nf}. 

Thus, in higher dimensions, to calculate correlators in the perturbed geometry in the geodesic approximation, we would need to use complexified geodesics as in \cite{Fidkowski:2003nf}. However, the shock wave spacetime is not an analytic solution, so it does not have a unique complex extension allowing us to calculate the lengths of these complex geodesics. This problem could perhaps be addressed by moving away from the shock wave approximation and modelling the effects of the perturbation as some smooth deformation, but this will lead to considerable technical complication, so we leave this for future work.

\acknowledgments

We are grateful for discussions with Yang Lei. AR is supported by an STFC studentship. SFR is supported in part by STFC under consolidated grant ST/L000407/1. 

\bibliographystyle{JHEP}
\bibliography{paper}

\end{document}